\documentclass[reprint,aps,amsmath,amssymb,twoside,prd,showkeys,superscriptaddress,twocolumn]{revtex4-1}
\usepackage{verbatim}
\usepackage{comment}
\usepackage{graphicx}
\usepackage[T1]{fontenc}
\usepackage{epsfig}
\usepackage{bm}
\usepackage{amssymb}
\usepackage{float}
\usepackage{amsmath}
\usepackage{subfigure}
\usepackage{dcolumn}
\usepackage{cancel}
\usepackage[colorlinks]{hyperref}
\usepackage[usenames,dvipsnames]{color}
\hypersetup{
     breaklinks=true,
    pdfstartview={FitH},    % fits the width of the page to the window
    colorlinks=true,       % false: boxed links; true: colored links
    linkcolor=blue,          % color of internal links
    citecolor=red,        % color of links to bibliography
    filecolor=magenta,      % color of file links
    urlcolor=blue,           % color of external links
    anchorcolor=green,      % Color for anchor text
    linktocpage=true
}

\def\doi{http://doi.org}

\newcommand{\HCd}{\mathcal{H}}

\def\HCdt0{\tilde{\HCd}_{0}}

\newcommand{\afffias}{Frankfurt Institute for Advanced Studies (FIAS), Ruth-Moufang-Strasse~1, 60438 Frankfurt am Main, Germany}
\newcommand{\affbgu}{Physics Department, Ben-Gurion University of the Negev, Beer-Sheva 84105, Israel}
\newcommand{\affbahamas}{Bahamas Advanced Study Institute and Conferences, 4A Ocean 
Heights, Hill View Circle, Stella Maris, Long Island, The Bahamas}

\begin{document}
\title{The Local Group as a test system for Modified Newtonian Dynamics}
\author{David Benisty}
\email{Corresponding Author: benidav@post.bgu.ac.il}
\affiliation{\affbgu}\affiliation{\afffias}
\author{Eduardo I. Guendelman}
\email{guendel@bgu.ac.il}
\affiliation{\affbgu}\affiliation{\afffias}\affiliation{\affbahamas}
\begin{abstract}
The Local Group (LG) is  {an appropriate} test system for Modified Newtonian Dynamics, since the acceleration of M31 galaxy is fully in the deep MOND regime $a \ll a_0$. We model the LG as a two body problem of $M31$ and the Milky Way (MW) galaxies.  {Extending previous studies, we also include the Cosmological Constant.} The assumption that in the big bang the galaxies emerged from the same place and approach to the measured distance and velocity today (the Timing Argument), predicts the total mass for the LG: $(0.447 \pm 0.005)\cdot 10^{12} M_{\odot}$. The corresponding motion of the LG predicts a past encounter. The ratio between the baryonic mass that MOND considers to the mass that Newtonian case predicted, which includes dark matter is $10.74 \pm 0.82$. This ratio agrees with the ratio between the dark matter and baryonic matter in other galaxies.
\end{abstract}
\maketitle
\section{Introduction}
Dark matter is one for the most profound unsolved phenomena in modern astrophysics and cosmology. The standard approach describes the dark matter as cold massive particles as Weakly Interacting Massive Particles (WIMPS) \cite{Tao:1989xn,Morales:2002ud,Iocco:2012jt,Conrad:2014tla,Rott:2012gh,Baudis:2013eba,DRUKIER:2013lva,daSilva:2014qba,Cui:2015eba,Arcadi:2017kky,Queiroz:2017kxt}, axions \cite{Peccei:1978fx,Davier:1987dv,Murayama:1998jb,Kim:1998sy,Kim:1999ia,vanBibber:2001ud,Geralis:2009ue,Kim:2009xp,Pajer:2013fsa} or very light axion-like particles \cite{Masso:2002ip,Galanti:2019sya,Ertas:2020xcc}.

The presence of dark matter in galaxies is observed from different measurements. The basic one is the mismatch between the predicted Keplerian velocity of orbiting stars in galaxies and the measured one \cite{Rubin:1980zd,Begeman:1991iy}. The measured velocity for large distances is approximately constant. In addition this constant velocity is related to the luminous mass through the Tully-Fisher relation \cite{Zwaan:1995uu,McGaugh:2000sr,TorresFlores:2011uc,McGaugh:2011ac,Chen:2019ftv}.

Modified Newtonian Dynamics is a different formulation that is capable to explain the flat rotation curves of galaxies \cite{Milgrom:1983zz,Milgrom:1983ca,Schee:2013wqa,2012IJMPD,Vagnozzi:2017ilo,Casalino:2018tcd}. MOND changes the Newton's Second Law (NSL) to:
\begin{equation}
F = m a \, \mu \left(\frac{a}{a_0}\right)
\end{equation}
where $\mu \left(\frac{a_0}{a}\right)$ is some function. For $a \gg a_0$ the function approaches one, $\mu \left(\frac{a}{a_0}\right) \rightarrow 1$, and produces the NSL. In the \text{deep}-MOND, $a \ll a_0$, the function approaches:
\begin{equation}
\mu \left(\frac{a}{a_0}\right) \rightarrow \frac{a}{a_0}
\end{equation}
In the deep-MOND regime, the function reduces to the linear approximation, $\mu \left(\frac{a}{a_0}\right) \rightarrow \frac{a}{a_0}$, which yields the modified NSL:
\begin{equation}
F = m \left(\frac{a^2}{a_0}\right).
\label{eq:deepMOND}
\end{equation}
This modified version gives the flat rotation curves of galaxies and the Tully-Fisher relation.  {This fit exists for single galaxies.} However we want to test the theory for a two galaxies system, where the relative accelerations is also in the deep-MOND regime. We find the Local Group (LG) of Galaxies as a good test system, since the estimated accelerations of the LG are in the deep-MOND regime as we will see.

 {Earlier estimations for the LG mass have been done with different methods \cite{Wang:2019ubx}: considering simulations
\cite{Li:2007eg,Gonzalez:2013pqa,Penarrubia:2014oda,Banik:2015nia,Carlesi:2016eud}, the Kahn-Woltjer Timing Argument (TA) as much as the virial theorem
\cite{1959ApJ130705K,vanderMarel:2012xp,vanderMarel2019,Chernin:2009ms,2009AA5071271C}, numerical action method \cite{Phelps:2013rra}, machine learning \cite{McLeod:2016bjk} and the disturbed Hubble flow \cite{2009AA5071271C}. The estimations predict that the mass is around $10^{12}$ solar masses ($M_{\odot}$).}

 {The LG approximately consist with two large galaxies: the Milky Way (MW) and Andromeda galaxy (M31). In the early universe the galaxies started from the same location and the current state of M31 is known from the latest measurements \cite{vanderMarel:2012xp,vanderMarel2019}. This picture known as the the Kahn-Woltjer Timing Argument (TA) and has been used to estimate the mass of the LG \cite{1959ApJ130705K}. In our analysis we compare between the predicted mass of the LG from the Newtonian case and the MONDian case. Because MOND is an alternative explanation for the Dark Matter, the predicted mass of LG in the MONDian case should predict the baryonic matter mass only. While the Newtonian case predicts the mass for the baryonic matter and the dark matter. We compare the ratio between the masses is the LG to other galaxy systems and see if the prediction of MOND yields a good approximation. }

The structure of this paper is as follows: Section \ref{sec:form} formulates the equation of motion for the LG dynamics, both in the Newtonian case and in the MONDian case. Section \ref{sec:dyn} discusses the LG dynamics and the Timing Argument. Section \ref{sec:mass} calculates the mass of the LG for the MONDian case.   {Section \ref{sec:lam} calculates the contribution of the Cosmological Constant for the mass.} Finally, section \ref{sec:dis} discusses the results. 

\section{Two body problem and MOND}
\label{sec:form}
The Cosmological Constant domination is considered to govern at  cosmological scales \cite{Chernin:2009ms,2009AA5071271C,Chernin:2015nna}. However, \cite{Eingorn:2012dg,Eingorn:2012jm,Partridge:2013dsa,Gonzalez:2013pqa,McLeod:2016bjk,McLeod:2019cfg} show that the Cosmological Constant effect in the LG scales ($1 Mpc$) changes the mass of LG to be $13 \%$ higher. Hence we include the Cosmological Constant contribution in our analysis. In General Relativity the effect on the motion of a test body can be considered in the framework of the spherically symmetric Schwarzschild vacuum solution with a Cosmological Constant background, which in the linearized approximation takes the form:
\begin{equation}
ds^2 =  (1+2\phi/c^2) (c dt)^2 - (1-2\phi/c^2) d\vec{x}^2
\end{equation}
where $c$ is the speed of light in vacuum, $\phi$ is the potential:
\begin{equation}
\phi = -\frac{G M}{r} - \frac{\Lambda c^2}{6}r^2,
\end{equation}
and $G$ is the Newtonian gravitational constant. The total Lagrangian for two particles in the center of mass system reads \cite{Jetzer:2006gn}:
\begin{equation}
\mathcal{L}/\mu = \frac{1}{2} v^2 + G\frac{M}{r} + \frac{\Lambda}{6} r^2,
\end{equation}
where $r$ is the relative distance, $v$ is the relative velocity, $M$ is the total mass and $\mu$ is reduced mass. In polar coordinate system $(r,\varphi)$ the relative distance variation reads \cite{Emelyanov:2015ina,Carrera:2006im}:
\begin{equation}\label{eq:ENL}
\ddot{r} = \frac{L^2}{r^3}-\frac{GM}{r^2} +  \frac{1}{3}\Lambda c^2 \, r,
\end{equation}
where $L$ is the conserved angular momentum per mass $L = r \, v_{\textit{tan}}$, and $v_{\textit{tan}}$ is the tangential velocity. There are two different contributions to the acceleration. One is the angular momentum term  $L^2/r^3$, and the other is the gravitational part with the Cosmological Constant, which is related to the generalized Newtonian force $g_N$:
\begin{equation}
g_N = -\frac{G M}{r^2} + \frac{1}{3}\Lambda c^2 r. 
\end{equation}
 {The total acceleration is obtained by using the Pythagorean Theorem. In the deep-MOND regime, the equation of motion reads:}
\begin{equation}
\ddot{r} - \frac{L^2}{r^3} = \text{Sign}(g_N) \sqrt{|g_N a_0|},
\end{equation}
In order to keep the direction of the modified acceleration as the direction of the Newtonian force, we introduce the "sign" of $g_N$ in the equation of motion.

Newtonian gravity allows us to transform to another accelerated frame which generates a uniform gravitational potential, which is compensated by a linear transformation of the Newtonian potential which leaves Newton's equations and Poisson equation invariant \cite{Benisty:2019wpm}. This symmetry is correlated to the translation symmetry of the Newtonian cosmology, where observing the Universe from another "center" corresponds to shifting to another accelerated frame, but where all the laws remain invariant. Such trivial formulations of MOND are not consistent with cosmology. However we explore the effects of MOND as a local formulation under the assumption that there exists a theory which does not violate cosmological principles on large scales and reduces to MOND in the appropriate limit.
\begin{figure}
 	\centering
\includegraphics[width=0.46\textwidth]{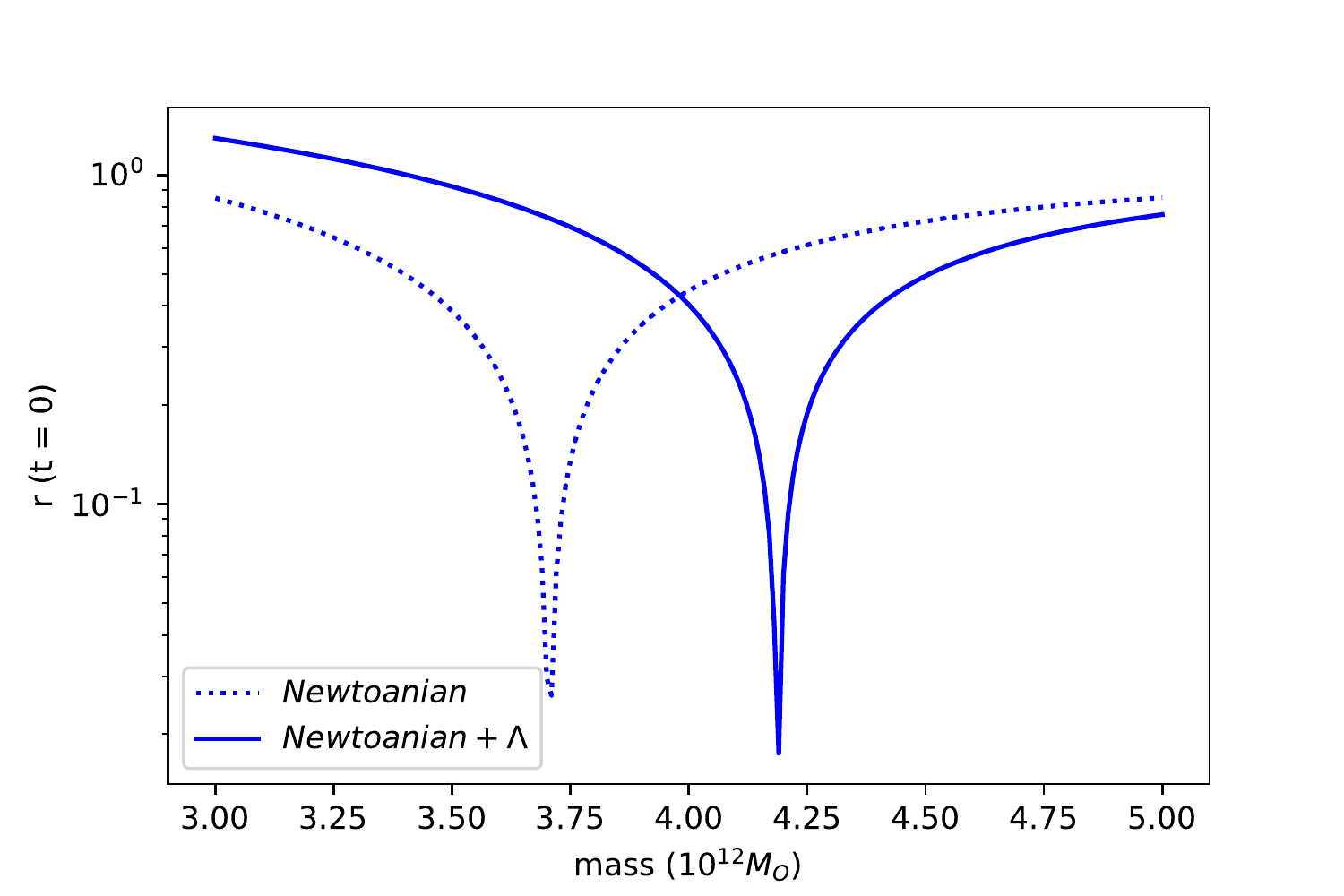}
\includegraphics[width=0.46\textwidth]{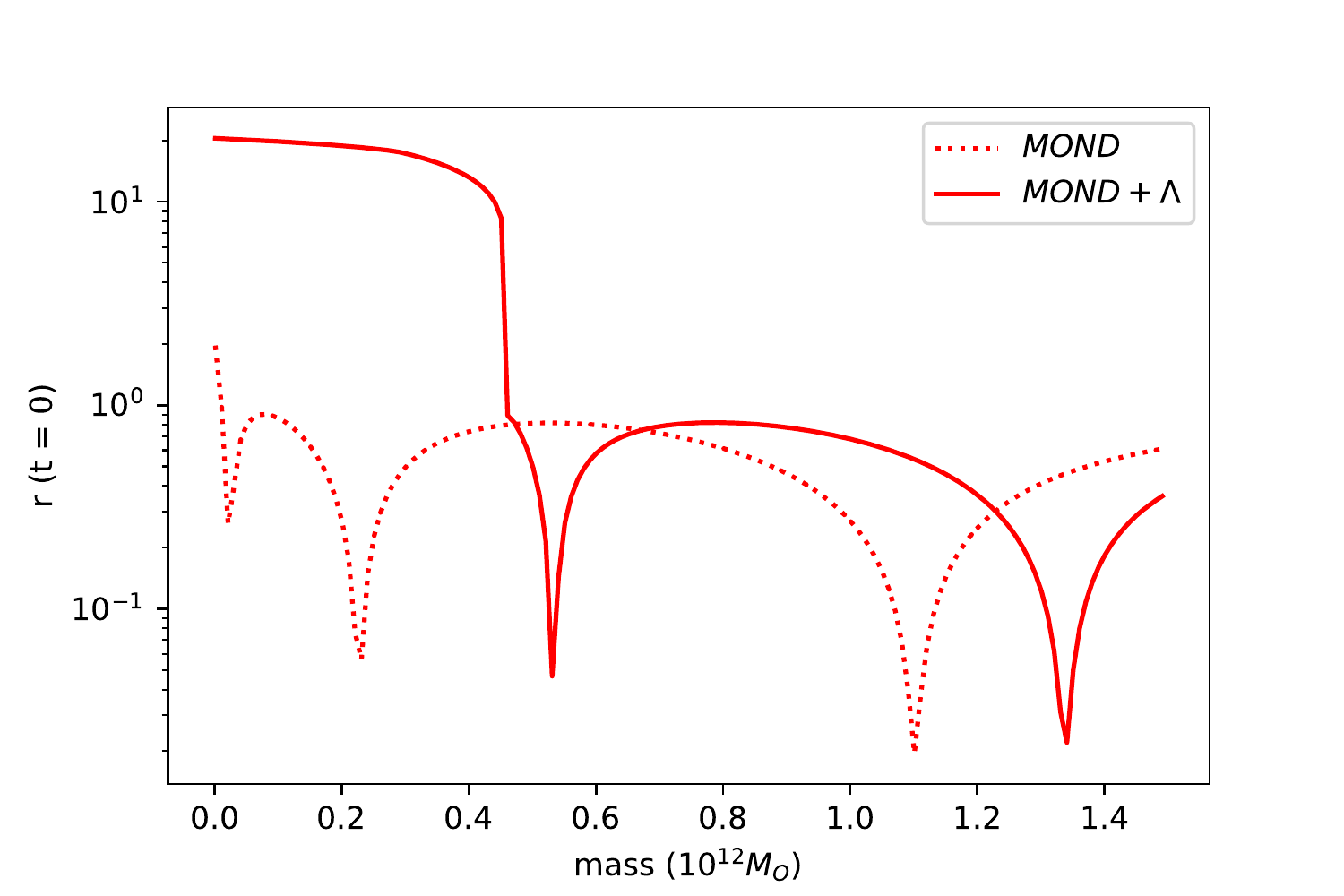}
\caption{\it{ {The distance at the big bang for different masses of the LG. The upper panel presents the Newtonian case, and the lower panel presents the Mondian case. The smooth and dashed lines refer to the cases with and without the Cosmological Constant, respectively. The prediction of the TA corresponds to a minimal point in the curve where $r$ approaches zero. Minimal points related to different number of past encounters.}}}
 	\label{fig:1}
\end{figure}

\begin{figure*}
 	\centering
\includegraphics[width=0.46\textwidth]{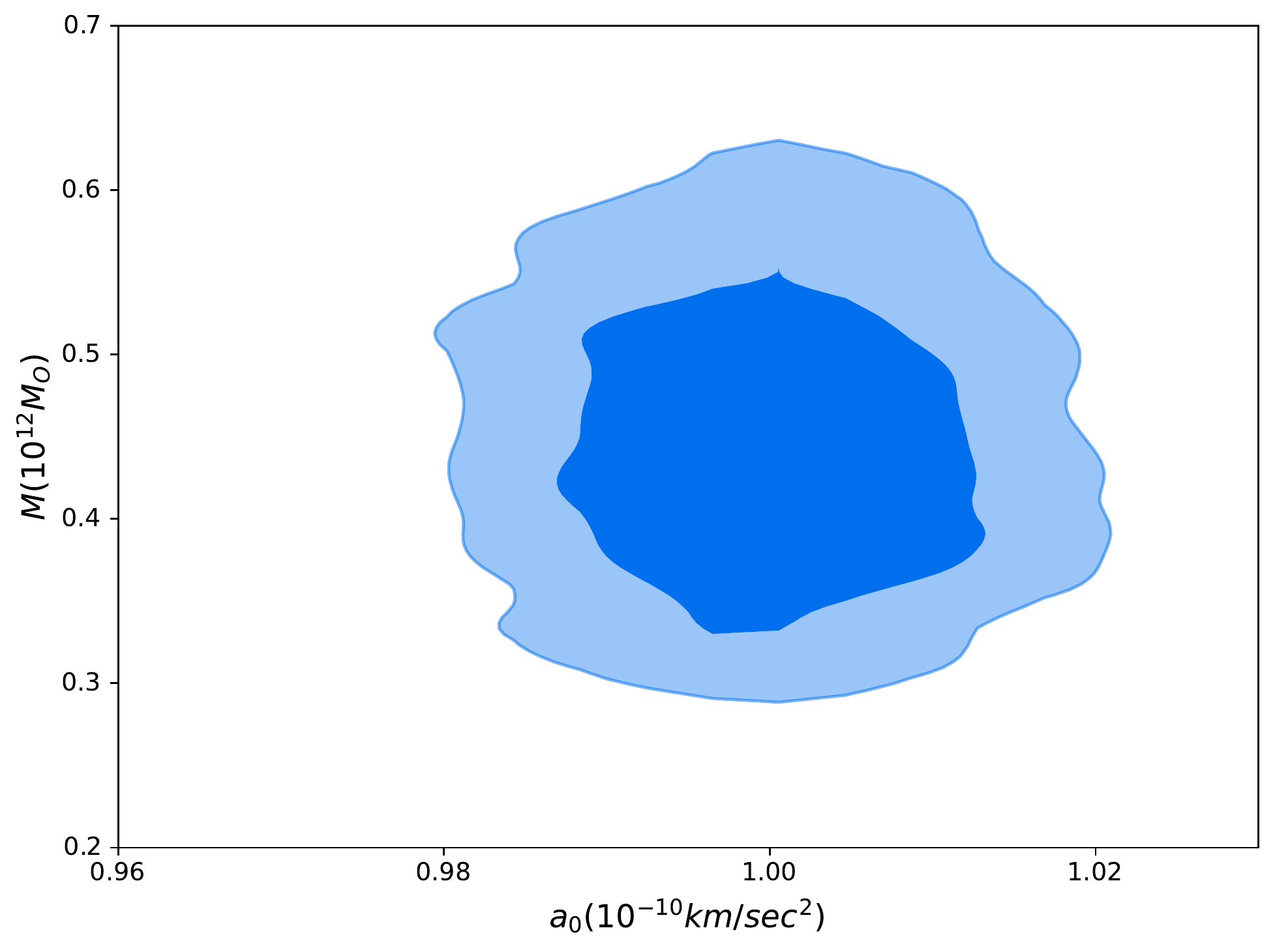}
\includegraphics[width=0.38\textwidth]{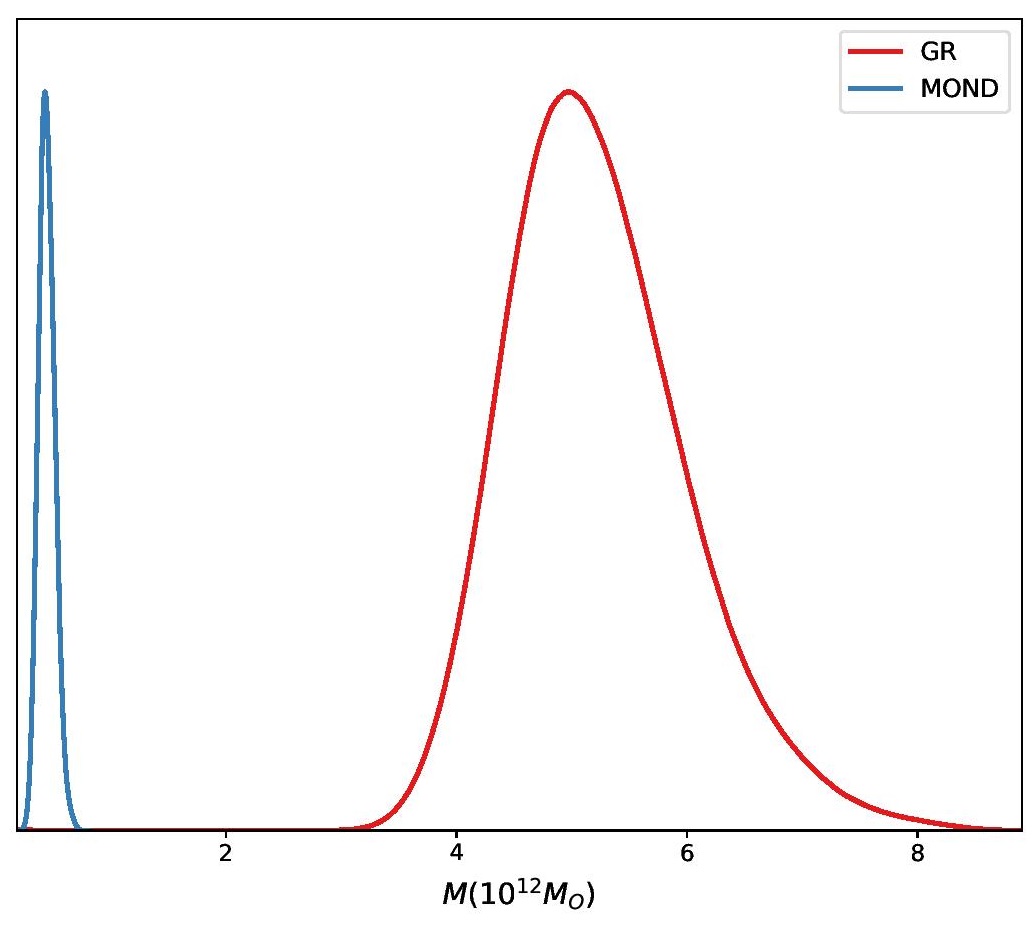}
\caption{\it{ {Left Panel: The contour plot for the total mass of the LG with a consideration of MOND with a Cosmological Constant background. Marginalizing over the parameters (r, $v_r$, $v_t$, $\Lambda_0$) yields the mass (\ref{eq:massMond}). Right Panel: The posterior distribution for the total mass of the LG for the Newtonian case and for the MONDian case. The mass of LG in Newtonian case is Eq. (\ref{eq:massGR}) and the mass of the MONDian case is Eq. (\ref{eq:massMond}).}}}
\label{fig:2}
\end{figure*}

\section{The LG Dimensions}
\label{sec:dyn}
\cite{vanderMarel:2012xp,vanderMarel2019} measure the final state of the M31 relatively to us is with the distance:
\begin{equation}
r_{m_{31}} = 0.77 \pm 0.04 \, {\text{Mpc}},
\label{eq:r}
\end{equation}
the radial velocity:
\begin{equation}
v_{rad} = -109.3 \pm 4.4 \, {km/sec},
\label{eq:vr}
\end{equation}
with the tangential velocity:
\begin{equation}
v_{tan} =  57^{+35}_{-31} \, {km/sec}.
\label{eq:vt}
\end{equation}
The Cosmological Constant is being:
\begin{equation}
\Lambda = (4.24 \pm 0.11) \cdot 10^{-66} \, eV^2
\label{eq:l}
\end{equation}
The age of the universe is:
\begin{equation}
 t_0 = 13.799\pm 0.021 \, Gys.
\label{eq:t0}
\end{equation}
Those values are determined also by the latest Planck measurements \cite{Aghanim:2018eyx}. The critical acceleration is taken to be:
\begin{equation}
a_0 \sim 10^{-10} \, {km/sec^2}    
\end{equation}
We can show that that the LG is in the deep MOND regime from the acceleration terms in Eq. (\ref{eq:ENL}). If we assume the total mass is around $10^{12} \, M_\odot$, the acceleration terms give:
\begin{equation}
\frac{GM}{r^2}\approx 2.06 \cdot 10^{-13} km/sec^2,
\end{equation}
\begin{equation}
\frac{1}{3}\Lambda c^2 \, r\approx 6.47 \cdot 10^{-14} km/sec^2,
\end{equation}
where both are in the deep-MOND regime. Therefore as we mentioned earlier, the LG is a good test system for MOND.

\section{The mass of the LG}
\label{sec:mass}
We evaluate the final state back in time. The galaxies are modeled as point masses. In order to calculate the mass of the Local group we evaluate the measured distance of $M31$ to obtain what should be the distance at the "big bang" for different LG masses. Fig.~\ref{fig:1} presents the distance for the big bang for different masses of LG. The model gives the predicted mass when the curve approaches minimum ($r \rightarrow 0$).

There are several minimal points in Fig~\ref{fig:1}. The blue line shows the distance at the big bang for the Newtonian case, while the red line shows the distance at the big bang for the MONDian case. The minimal point of the red line corresponds to the predicted mass for the MONDian case with one past encounter. Because the galaxies are not point like masses, the prediction seems to be true only for the first minimal point. Higher encounters cause the galaxies to merge.

Because of the distribution of the measured final conditions, we use a Gaussian prior for the initial distance (\ref{eq:r}), the radial velocity (\ref{eq:vr}), the tangential velocity (\ref{eq:vt}), the Cosmological Constant (\ref{eq:l}) and the age of the universe (\ref{eq:t0}) similarly to \cite{Benisty:2019fzt}, where the error bar of the initial condition is taken to be the variance. However, for $a_0$ we use a uniform prior of $ a_0 \in [0.1;10]\cdot10^{-10} \, km/sec^2$. We use Monte Carlo simulation with $10^7$ samples.

\begin{figure}
 	\centering
\includegraphics[width=0.44\textwidth]{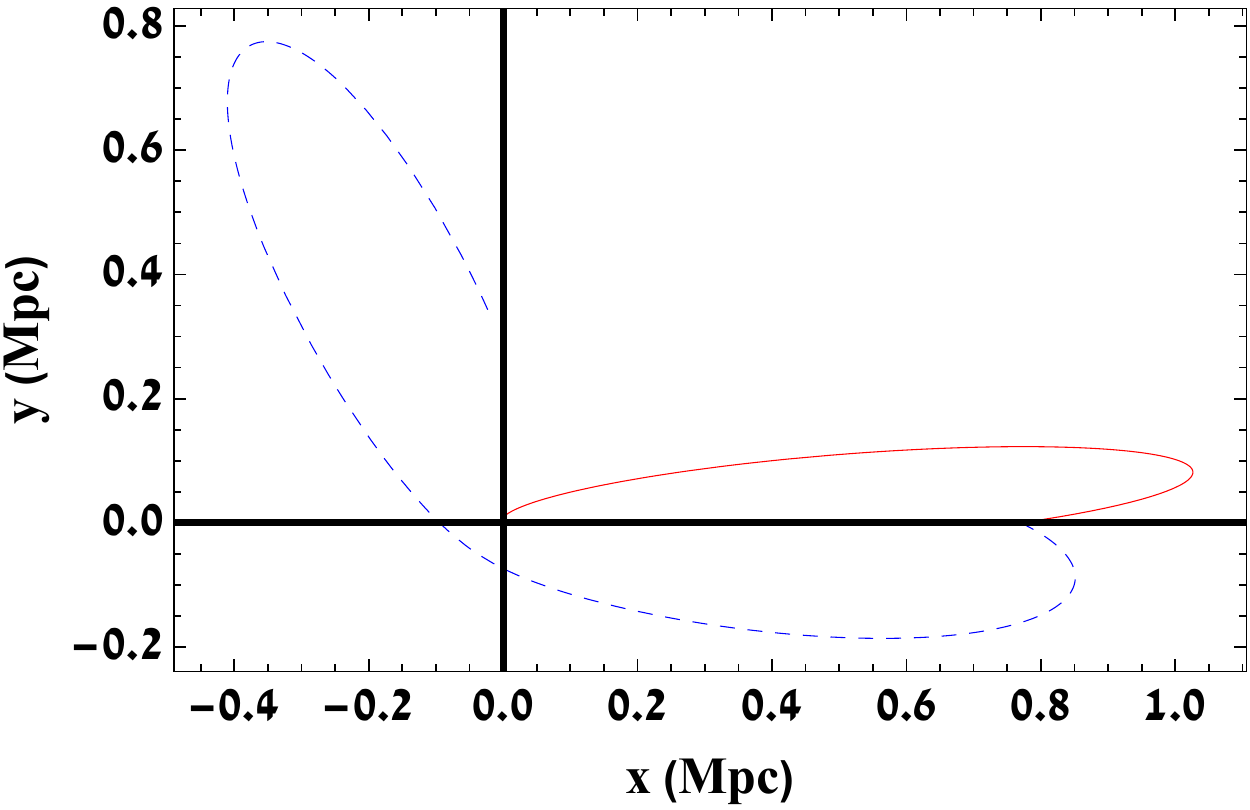}
\caption{\it{The predicted motion of M31 from the big bang  (at the origin) up to the final location (0.77 Mpc,0). The red line refers to the Newtonian case and the blue dashed line refers to the MONDian case with one past encounter.}}
 	\label{fig:3}
\end{figure} 
\begin{figure}
 	\centering
\includegraphics[width=0.48\textwidth]{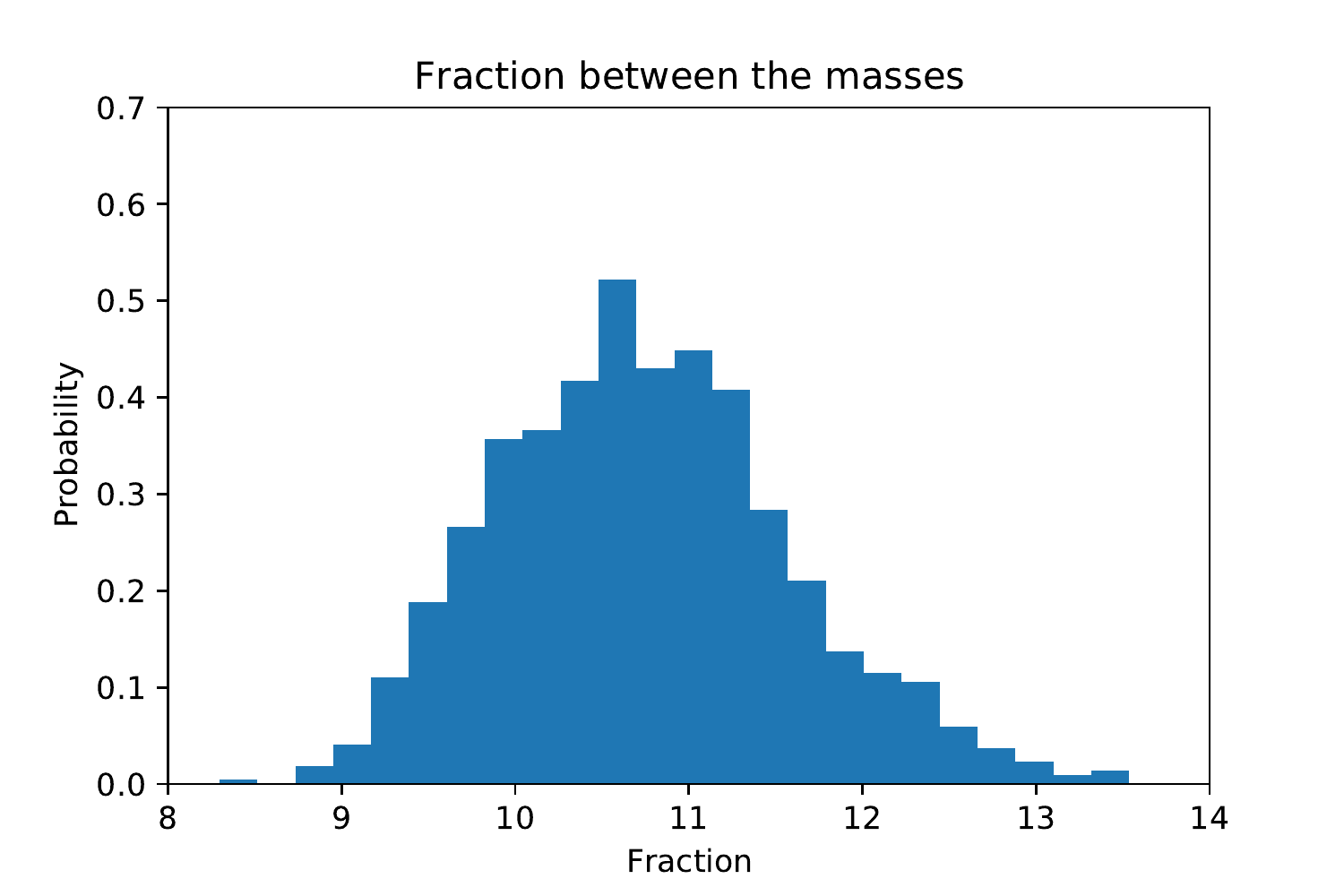}
\caption{\it{The statistical distribution for the ratio between the predicted dark matter and the predicted baryonic matter. The distribution yields the ratio $\approx10$.}}
 	\label{fig:4}
\end{figure}

Fig \ref{fig:2} shows the posterior distribution for the mass with a consideration of MOND vs. the value of $a_0$, with a Cosmological Constant. The mass of LG in Newtonian case, as observed originally in \cite{Benisty:2019fzt}, is being: 
\begin{equation}
M_\text{Newtonian} = \left(5.23 \pm 0.63\right) 10^{12} M_{\odot}.
\label{eq:massGR} 
\end{equation}
The mass for the MONDian case is:
\begin{equation}
M_\text{MOND} = \left( 0.447 \pm 0.005 \right) 10^{12} M_{\odot}. \label{eq:massMond}     
\end{equation}
 \begin{figure*}
 	\centering
\includegraphics[width=0.9\textwidth]{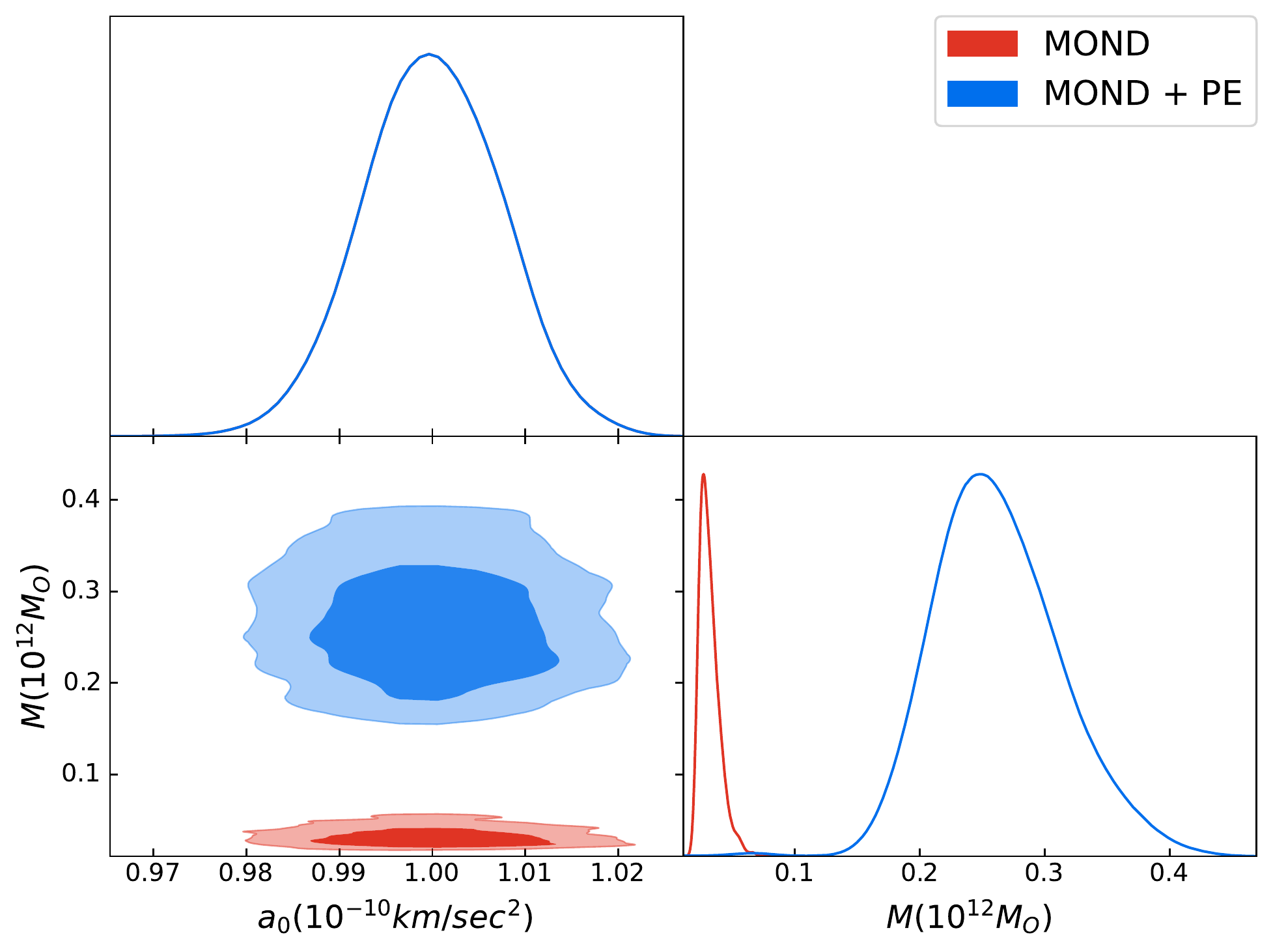}
\caption{\it{ {The contour plot for the total mass of the LG with a consideration of MOND without a Cosmological Constant, marginalizing over the parameters (r, $v_r$, $v_t$) yields the mass (\ref{eq:massMondNoLam}) and (\ref{eq:massMondNoLamPE}) for one past encounter.}}}
 	\label{fig:5}
\end{figure*} 

In order to track the actual motion of M31 in both gravity models, we integrate the $\dot{\varphi}$ and calculate the angle $\varphi = \int L/r^2 dt$. The final conditions for the numerical solution is $\varphi = 0$. Fig \ref{fig:3} shows the actual motion of M31 relatively to the MW galaxy. The red line corresponds to the Newtonian prediction and the dashed blue line corresponds to the MONDian prediction. In both cases the M31 galaxy begins at the origin and finish at the position $(0.77_{ Mpc},0)$. The Newtonian case describes how M31 getting away and getting closer. But the MONDian case predicts one past encounter. \cite{Zhao:2013uya} already predicted this past encounter. However, \cite{Zhao:2013uya} didn't take into account the Cosmological Constant in addition the modified inertia of MOND.

Past encounters are problematic in the Newtonian case due to the strong Dynamical Friction (DF) from the dark matter \cite{1987gady,Hammer:2007ki}. \cite{Cox:2007nt,Conselice} uses an N-body simulation to track the evolution of the LG, focusing on the Milky Way and Andromeda. The simulation shows that DF between the galaxies would lead to the eventual merger of the LG. However, for MOND the scenario is different: it provides an alternative law of inertia and therefore there is less matter and less DF. \cite{2018MNRAS4734033B,2018A&A614A59B} claim from N-body simulation that in MOND the galaxies would not merge after the past encounter.
 
The ratio between the masses requires from  us a different analysis for MOND. MOND is a formulation that replaces dark matter. So the MONDian TA gives a prediction that all the mass is the baryonic matter alone. The Newtonian case predicts the baryonic and an additional amount of  dark matter constitute the total mass.  The dark matter mass is calculated by the difference between the Newtonian prediction and MONDian prediction. Fig \ref{fig:4} shows the distribution for the ratio between the dark matter and the baryonic matter. The distribution yields the ratio:
\begin{equation}
M_{DM}/M_{b} = 10.74 \pm 0.82,   
\end{equation}
with $1\sigma$ error. The ratio between the dark matter and baryonic matter in our universe is around six, where for different galaxies the ratio is approximately ten \cite{Edmonds:2013hba,Posti:2018vts}.  {The ratio agrees with the ratio in  some galaxies.}

\section{The contribution of $\Lambda$}
\label{sec:lam}
 {In order to complete our analysis, we test the predicted mass of the LG without the presence of the Cosmological Constant. Marginalizing over the initial conditions yields the mass:}
\begin{equation}
M_{MOND} (\Lambda = 0) = \left( 3.16 \pm 0.07 \right) 10^{10} M_{\odot}. 
\label{eq:massMondNoLam}
\end{equation}
This value agrees with earlier estimations, such as: $0.027 \cdot 10^{12} M_{\odot}$, that \cite{McLeod:2019cfg} predicted. This value does not corresponds to a past encounter, but the motion begins in the big bang ($r = 0$) and ends with the measured relative distance today. Fig \ref{fig:5} shows the contour plot for the mass (red curve). In addition to first minimum for the dashed line in Fig \ref{fig:1}, there is a second minimum that predicts a possible mass. This minimum corresponds to the mass:
\begin{equation}
M_{MOND, PE} (\Lambda = 0) = \left( 2.62 \pm 0.03 \right) 10^{11} M_{\odot}. 
\label{eq:massMondNoLamPE}
\end{equation}
Fig \ref{fig:5} shows the contour plot for the mass (blue curve) for the case of one past encounter. The mass agrees with the prediction for the case with the Cosmological Constant. Because the Cosmological Constant pushes the galaxies against the gravitational force, its presence results in a larger mass in order to fit the initial condition of the TA and the measured final conditions. Notice also that the tangential velocity of M31, measured by \cite{vanderMarel2019}, is larger then the earlier estimations. So we expect for slightly different predicted masses.

 {Our analysis predicts that one PE for the case with $\Lambda$, is an essential solution, since there is no a minimal point in Fig \ref{fig:1} that yields a solution without PE. For the case of zero $\Lambda$ the minimal point that predicts one PE is different from the first minimal point, that does not yield a past encounter. }

\section{Discussion}
\label{sec:dis}

This paper we test the mass of LG in the MOND formulation. we are treating the MW and M31 galaxies as point particles that emerge briefly after the big bang at a very small distance. The requirement that M31 has at present the distance and velocity as observed allows one to extract the mass of LG. %In order to have the same final conditions for the M31 distance and velocity at the present time, one can extract the mass for the LG. 

 {MOND is a formulation that modifies the Newtonian Second Law for low accelerations instead of dark mater. The ratio between the dark matter and baryonic matter according to the $\Lambda$CDM model is around six, where for different galaxies the ratio is approximately ten and higher  \cite{Edmonds:2013hba,Posti:2018vts} . Therefore, if the Newtonian TA predicts that the mass of LG should be around $\left(5.2\pm 1.7\right) 10^{12} M_{\odot}$, then the baryonic matter should be around $\left(0.45\pm 0.15 \right)\cdot 10^{12} M_{\odot}$ for the MONDian cases. The MONDian case forces one past encounter. }

 {Notice that the deep MOND approximation will be correct most of the time, except very close to the encounter, where Newtonian dynamics becomes valid again due
to the larger accelerations. This does not affect however the validity of the conclusions concerning encounters since this is a very small section of the trajectories
of the galaxies. At the encounter itself we will have now a Newtonian situation but with much smaller masses, so may be large dynamical friction effects could be avoided and also
the definite merger of the two galaxies, which of course will be against the observed two separate galaxies that we see at the present time for our LG.}

Finally, it will be important to extend the analysis to modified theories of gravity that predict a linear component in the gravitational potential instead of modifying the inertia as MOND. This model could be useful to test the validity for those theories, whenever the predicted mass would be much smaller than the Newtonian model. These models could arise from conformal gravity \cite{Mannheim:1992vj} or alternatives theories of gravity \cite{Bahamonde:2018uao,Nojiri:2010wj,Berti:2015itd,Nojiri:2017ncd,Vagnozzi:2019kvw,DiValentino:2020evt}.

\acknowledgments
We wish to thank Doug Edmonds for engaging and very important discussions concerning MOND theory. We thank Ofer Lahav, Mordehai Milgrom, Indranil Banik and Hongsheng Zhao for stimulating discussions. This article is supported by COST Action CA15117 "Cosmology and Astrophysics Network for Theoretical Advances and Training Action" (CANTATA) of the COST (European Cooperation in Science and Technology). We have received partial support from European COST action CA18108 "Quantum gravity phenomenology in the multi-messenger approach". D.B. thankful to Bulgarian National Science Fund for support via research grant KP-06-N 8/11.

\bibliography{ref}
\bibliographystyle{apsrev4-1}

\end{document}